\def\astroph{1}

\ifnum\astroph=0
\documentclass[12pt,preprint]{aastex}       

\else
\documentclass{emulateapj}
\usepackage{apjfonts}
\fi

\usepackage{epsfig,amsmath,natbib}

\begin{document}

\shorttitle{Collisionless Shocks} \shortauthors{Hededal and
Nishikawa}

\title{The Influence of an Ambient Magnetic Field on Relativistic Collisionless Plasma Shocks}
\author{Christian Busk Hededal and Ken-Ichi Nishikawa}
\affil{Niels Bohr Institute, Deptartment of Astrophysics, Juliane
Maries Vej 30, 2100 K\o benhavn \O , Denmark; and National Space
Science and Technology Center (NSSTC), Gamma-Ray Astrophysics Team,
320 Sparkman Drive, Huntsville, AL 35805;}
\email{hededal@astro.ku.dk}

{\it Received 2004 December 13; accepted 2005 March 15; published
2005 April}

\begin{abstract}
Plasma outflows from gamma-ray bursts, supernovae, and relativistic
jets, in general, interact with the surrounding medium through
collisionless shocks. The microphysics of such shocks are still
poorly understood, which, potentially, can introduce uncertainties
in the interpretation of observations. It is now well established
that the Weibel two-stream instability is capable of generating
strong electromagnetic fields in the transition region between the
jet and the ambient plasma. However, the parameter space of
collisionless shocks is vast and still remains unexplored. In this
Letter, we focus on how an ambient magnetic field affects the
evolution of the electron Weibel instability and the associated
shock. Using a particle-in-cell code, we have performed
three-dimensional numerical experiments on such shocks. We compare
simulations in which a jet is injected into an unmagnetized plasma
with simulations in which the jet is injected into a plasma with an
ambient magnetic field both parallel and perpendicular to the jet
flow. We find that there exists a threshold of the magnetic field
strength below which the Weibel two-stream instability dominates,
and we note that the interstellar medium magnetic field strength
lies well below this value. In the case of a strong magnetic field
parallel to the jet, the Weibel instability is quenched. In the
strong perpendicular case, ambient and jet electrons are strongly
accelerated because of the charge separation between deflected jet
electrons and less deflected jet ions. Also, the electromagnetic
topologies become highly non-linear and complex with the appearance
of anti-parallel field configurations.
\end{abstract}


\keywords{acceleration of particles,
gamma rays: bursts, shock waves, instabilities, magnetic fields,
plasmas}

\section{Introduction}
The collisionless plasma condition applies to many astrophysical
scenarios, including, the outflow from gamma-ray bursts (GRBs),
active galactic nuclei, and relativistic jets in general. The
complexity of kinetic effects and instabilities makes it difficult
to understand the nature of collisionless shocks. Only recently, the
increase in available computational power has made it possible to
investigate the full three-dimensional dynamics of collisionless
shocks.

In the context of GRB afterglows, observations indicate that
shock-compressed magnetic field from the interstellar medium (ISM)
is several orders of magnitude too weak to match observations.
Particle-in-cell (PIC) simulations have revealed that the Weibel
two-stream instability is capable of generating the required
electromagnetic field strength of the order of percents of
equipartition value (\cite{bib:kazimura1998, bib:medvedevloeb,
bib:nishikawa, bib:nishikawa2005,  bib:silva, bib:frederiksen}).
Furthermore, PIC simulations have shown that in situ non thermal
particle acceleration takes place in the shock transition region
\citep{bib:hoshino2002, bib:hededal2004, bib:saito2003}.
Three-dimensional simulations using $\sim 10^7$ electron-positron
pairs by \cite{bib:sakai} showed how complex magnetic topologies are
formed when injecting a mildly relativistic jet into a force-free
magnetic field with both parallel and perpendicular components. With
a two-dimensional analysis, \cite{bib:saito2003} found that an
ambient parallel magnetic field can quench the two-stream
instability in the weakly relativistic case. In this Letter, we use
three-dimensional PIC experiments to investigate how the two-stream
instability is affected by the presence of an ambient magnetic
field. Using up to $\sim10^9$ particles and $125\times125\times1200$
grid zones, we investigate the development of complex magnetic
topologies when injecting a fully relativistic jet (bulk Lorentz
factor $\Gamma=5$) into an ambient magnetized plasma. Using varying
field strengths, we focus on the case of a transverse magnetic field
and compare it with the case of a parallel magnetic field. The
simulations are mainly concerned with the electron dynamics since
processes involving the heavier ions evolve on much longer
timescales.

\section{The numerical experiments}
\begin{figure}[!ht]
\begin{center}
\epsfig{figure=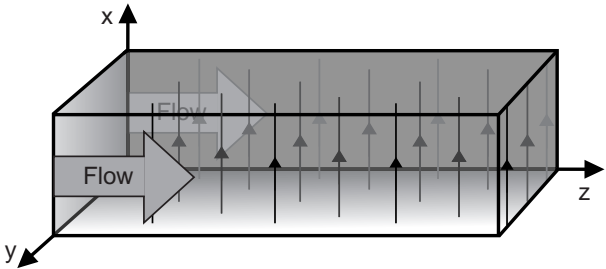,width=8cm} \caption{Schematic example of the
simulation setup. Jet plasma is homogeneously and continuously
injected in the $z$-direction throughout the $x-y$ plane at $z=0$.
Inside, the box is populated by a plasma population, initially at
rest. In this specific example, an ambient magnetic field is set up
in the $x$-direction (perpendicular case).} \label{fig:setup}
\end{center}
\end{figure}

We use the PIC code described by \cite{bib:frederiksen}. The code
works from first principles and evolves the equation of motion for
the particles together with Maxwell's equations.
In the simulation experiments, we inject an electron-proton plasma
(a jet) into an ambient plasma (the ISM) initially at rest (Fig.
\ref{fig:setup}). The jet is moving with a relativistic velocity of
$0.98c$ along the $z$-direction corresponding to Lorentz factor
$\gamma_{jet}=5$. The ion-to-electron mass ratio is set to $m_{\rm
i}/m_{\rm e}=20$. The jet plasma and the ambient plasma have the
same density, $n$, and the corresponding electron plasma rest-frame
frequency $\omega_{pe}\equiv [ne^2/(m_e\epsilon_0)]
=0.035\Delta_t^{-1}$ ($e$ is the unit charge, $\epsilon_0$ the
vacuum permittivity, and $\Delta_t$ the simulation unit time). We
choose this low value in order to properly resolve the microphysics.
Initially, the interface between the ambient and the injected plasma
is located at $z=3\lambda_e$, where $\lambda_e$ is the electron skin
depth defined as $\lambda_e\equiv c/\omega_{pe}=28.6\Delta_x$ ($c$
is the speed of light, and $\Delta_x$ the grid size). The time step
and grid size obey the Courant-Friedrichs-Levy condition
$\Delta_t=0.5\Delta_x/c$. Both plasma populations are, in their
respective rest frames, Maxwellian distributed with a thermal
electron velocity $v_{th}\simeq0.03c$. This temperature allow us to
numerically resolve the plasma Debye length with approximately one
grid length.

We consider three different ambient magnetic configurations: no
magnetic field, a magnetic field parallel to the flow, and a
magnetic field perpendicular to the flow. The magnetic field is
initially setup to be homogeneous and at rest in the ambient plasma.
The experiments are carried out with 1 billion particles inside
$(125\times125\times1200)$ grid zones. In terms of electron skin
depths, this corresponds to $(4.4, 4.4, 42) \lambda_e$. The boundary
conditions are periodic in the direction transverse to the jet flow
($x, y$). In the parallel direction, jet particles are continuously
injected at the leftmost boundary ($z=0$). At the leftmost and
rightmost $z$ boundary, electromagnetic waves are absorbed, and we
allow particles to escape in order to avoid unphysical feedback. The
total energy throughout the simulations is conserved with an error
less than $1\%$.

\section{Results}
Initially, we run simulations with no ambient magnetic field and
observe the growth of the Weibel two-stream instability also found
in previous work (\cite{bib:kazimura1998, bib:medvedevloeb,
bib:nishikawa,bib:nishikawa2005, bib:silva, bib:frederiksen}). The
Weibel two-stream instability works when magnetic perturbations
transverse to the flow collect streaming particles into current
bundles that in turn amplify the magnetic perturbations. In the non
linear stage, we observe how current filaments merge into
increasingly larger patterns. The electromagnetic energy grows to
$\epsilon_B\simeq1\%$, where $\epsilon_B$ describes the amount of
total injected kinetic energy that is converted to magnetic energy.

\subsection{Parallel Magnetic Field}
\begin{figure}[!ht]
\begin{center}
\epsfig{figure=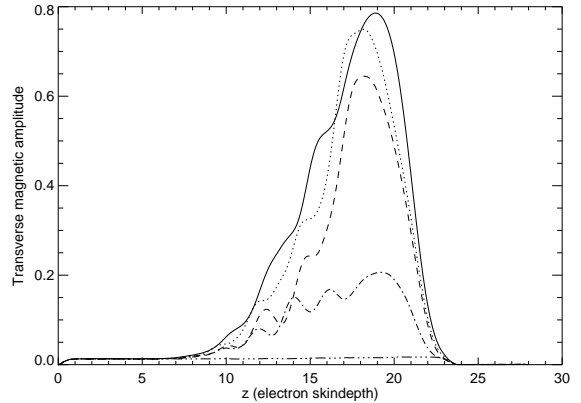,width=8cm} \caption{ Growth of the Weibel
two-stream instability for different strengths of the parallel
ambient magnetic field at time $t = 21\omega_{pe}^{-1}$. Here we
measure the effectiveness of the field generation as the average
transverse magnetic field amplitude as a function of $z$. The solid
line corresponds to $\omega_{pe}/\omega_{ce}=40$,
 the dotted line to 20, the dashed line to 10, the dot-dashed line
to 5, and the triple-dot-dashed line to 1. The case of no ambient
magnetic field is very similar to that of
$\omega_{pe}/\omega_{ce}=40$. The magnetic amplitude is in arbitrary
units} \label{fig:b_par}
\end{center}
\end{figure}

In the presence of a strong magnetic field component parallel to the
flow, particles are not able to collect into bundles since
transverse velocity components are deflected. We have performed five
runs with parallel magnetic fields corresponding, respectively, to
$\omega_{pe}/\omega_{ce}= $40, 20, 10, 5, and 1 while keeping
$\omega_{pe}$ constant; $\omega_{ce}=eB/(\gamma_{jet} m_e)$ is the
jet electron gyrofrequency. The resulting field generation
efficiency can be seen in Fig. \ref{fig:b_par} at $t =
21\omega_{pe}^{-1}$ where the jet front has reached $z =
23\lambda_{e}$. In the case of $\omega_{pe}/\omega_{ce}=40$, the
Weibel instability overcomes the parallel field, and although
initially slightly suppressed, it eventually evolves as in the case
of no ambient magnetic field. Increasing the magnetic field to
$\omega_{pe}/\omega_{ce}=1$ effectively suppresses the instability.
Thus, for an ISM strength magnetic field
($\omega_{pe}/\omega_{ce}\simeq1500$) parallel to the plasma flow,
the Weibel two-stream instability will evolve unhindered, and the
generated field will exceed the ISM field. We find from the
simulations that it would take a milligauss strength parallel
magnetic field to effectively quench the instability for a
$\gamma=5$ jet expanding in an environment with density like the
ISM.

\begin{figure*}[!ht]
\begin{center}
\epsfig{figure=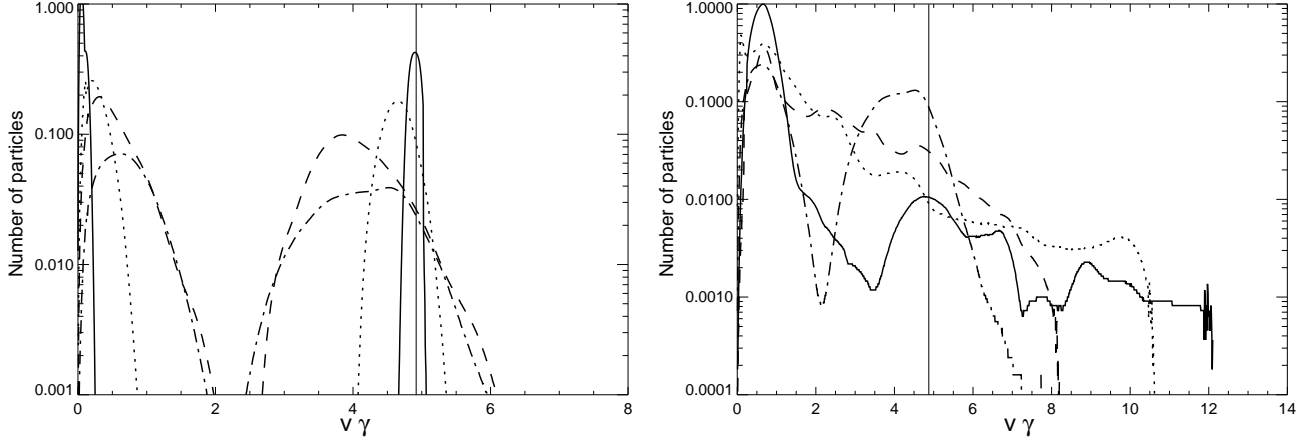,width=17.5cm} \caption{Normalized electron
momentum distribution functions at time $30 \omega_{pe}^{-1}$. The
left panel is for runs with an ambient magnetic field parallel to
the injected plasma with $\omega_{pe}/\omega_{ce}=$ 1 ({\em solid
line}), 20 ({\em dotted line}), 40 ({\em dashed line}), and 1500
({\em dot-dashed line}). The right panel is for runs in which the
initial magnetic field is perpendicular to the inflow and
$\omega_{pe}/\omega_{ce}=$ 5 ({\em solid line}), 20 ({\em dotted
line}), 40 ({\em dashed line}) and 1500 ({\em dot-dashed line}). The
vertical line shows the injected momentum $\gamma=5$. The
distribution functions are for electrons with $z>15\lambda_e$.}
\label{fig:pdf}
\end{center}
\end{figure*}

The left panel of Fig. \ref{fig:pdf} show the resulting electron momentum
distribution function for different values of $\omega_{pe}/\omega_{ce}$
 at
$t=30\omega_{pe}^{-1}$. Since the presence of a strong parallel
magnetic field suppresses the generation of a transverse magnetic
field, there exists no mechanism that can heat the electrons and
transfer momentum between the two electron populations. Thus the jet
plasma propagates unperturbed. Where there is no parallel magnetic
field or only a weak magnetic field
($\omega_{pe}/\omega_{ce}=1500$), we observe how the jet and ambient
plasma is heated and how momentum is transferred between the two
populations.

\subsection{Perpendicular Magnetic Field}
We have performed experiments with an ambient magnetic field
perpendicular to the jet flow (Fig. \ref{fig:setup}) with field
strengths corresponding to $\omega_{pe}/\omega_{ce}$=1500, 40, 20
and 5. By including the displacement current, one can derive the
relativistic Alfv\'en speed
$v_A^{-2}=c^{-2}+(v_{A}^{non-rel.})^{-2}\simeq
c/[1+(\omega_{pe}/\omega_{ce})^2 (m_i/m_e)\gamma_{jet}^{-2}]^{1/2}$,
where $v_{A}^{non-rel.}=B/[\mu_0 n (m_i+m_e)]^{1/2}$ is the
non-relativistic counterpart. From this we calculate the
corresponding relativistic Alf\'ven Mach numbers
$\gamma_{jet}v_{jet}/v_A=$ 6572, 175, 88, and 22.

Again, the $\omega_{pe}/\omega_{ce}=1500$ run has been chosen
because it resembles the typical density and microgauss magnetic
field strength of the ISM. We find that the magnetic field generated
by the two-stream instability dominates the ambient magnetic field,
and the result resembles that of no ambient magnetic field.
Furthermore, as expected in both the parallel and perpendicular
cases, the electron momentum distributions (Fig. \ref{fig:pdf}) are
very similar, except for a weak merging between the ambient and jet
electrons in the perpendicular case.
\begin{figure}[!ht]
\begin{center}
\epsfig{figure=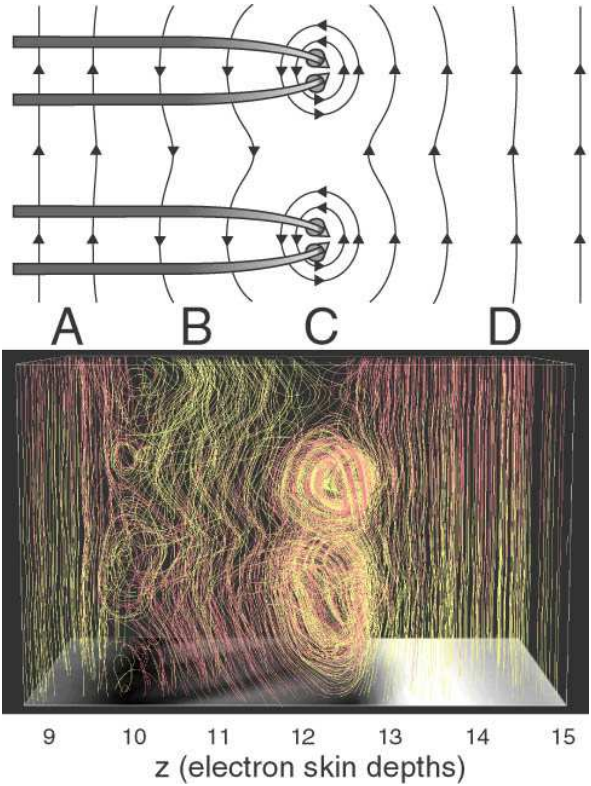,width=7.8cm} \caption{Snapshot at
$t=16\omega_{pe}^{-1}$ of the highly complicated magnetic field
topology when a jet is injected into a plasma with an ambient
magnetic field transverse to the jet flow. The bottom panel shows
magnetic field lines in a subsection of the computational box (from
$z=9\lambda_e$ to $z=15\lambda_e$. The top panel shows a schematic
explanation in the $x-z$ plane: Jet electrons are bent by the
ambient magnetic field (region A). Due to the Weibel instability,
the electrons bundle into current beams (region C) that in turn
reverse the field topology (region B). This will eventually bend the
jet beam in the opposite direction. [{\it See the electronic edition
of the Journal for a color version of this figure.}]}
\label{fig:step1}
\end{center}
\end{figure}

\begin{figure}[!h]
\begin{center}
\epsfig{figure=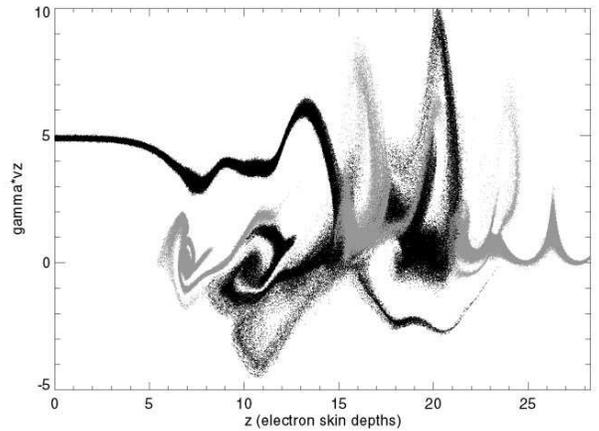,width=8.5cm} \caption{Electrons $v_z\gamma$
plotted against their position in the shock at time equal
$30\omega_{pe}^{-1}$ for the run with an ambient magnetic field
transverse to the jet flow corresponding to
$\omega_{pe}/\omega_{ce}=20$. The jet electrons ({\em black dots})
are injected at $z=0$ with $\gamma=5$. The ambient electrons ({\em
gray dots}) are initially at rest but are strongly accelerated by
the jet.} \label{fig:scatter}
\end{center}
\end{figure}

In the run with $\omega_{pe}/\omega_{ce}=20$, the result differs
substantially from the previous cases. With reference to Fig.
\ref{fig:step1}, we describe the different stages of the evolution:
Initially, the injected particles are deflected by the ambient
magnetic field. The magnetic field is piled up behind the jet front,
and the enhanced magnetic fields bend jet electron trajectories
sharply. This has two implications. First, the ions, being more
massive, will penetrate deeper than the deflected electrons. This
creates a charge separation near the jet head that effectively
accelerates both ambient and injected electrons behind the ion jet
front as shown in Fig. \ref{fig:scatter}. Second, the deflected
electrons eventually become subject to the Weibel two-stream
instability. This forms electron current channels at some angle to
the initial direction of injection, as shown in the upper panel of
Fig. \ref{fig:step1}. Around these current channels, magnetic loops
are induced (Fig. \ref{fig:step1}, region C). Magnetic islands are
formed and the ambient magnetic field is reversed behind the loops
(region B). In this region, we find acceleration of electrons in the
$x$-direction. The activity in this region has similarities to
reconnection, but it is beyond the scope of this Letter to
investigate this topic.

In other regions, the ambient magnetic field is strongly compressed,
and this amplifies the field strength up to 5 times the initial
value. As a result, parts of the jet electrons are actually reversed
in their direction. This can be seen from Fig. \ref{fig:scatter},
which shows a phase-space plot of both ambient and jet electrons at
$t = 30\omega_{pe}^{-1}$. We see several interesting features here.
In the region $z=(15-21)\lambda_e$, we observe how ambient electrons
are swept up by the jet. Behind the jet front, both ambient and jet
electrons are strongly accelerated since the jet ions, being
heavier, take a straighter path than the jet electrons, and this
creates a strong charge separation. The excess of positive charge in
the very front of the jet head is very persistent and hard to shield
since the jet ions are moving close to the speed of light. Thus,
there is a continuous transfer of $z$-momentum from the jet ions to
the electrons. In the case of the perpendicular ambient field, more
violent processes take place than in the case of the parallel field,
which can be seen in Fig. \ref{fig:pdf} ({\it right panel}). Here we
see that mixing of the two plasma populations is much more effective
for the perpendicular case. However, the spectrum of the electrons'
momentum is highly nonthermal, with strong acceleration of both jet
and ambient electrons. The cutoff in electron acceleration depends
on the magnetic field strength. The maxima ($\gamma v_{\parallel}
\approx 10$) at $z = 20\lambda_{e}$ in Fig \ref{fig:scatter}
corresponds to the cutoff shown by the dotted line in the right
panel in Fig. \ref{fig:pdf}. It should be noted that the current
channels that are caused by the bent jet electron
 trajectories
in the early time, as shown in Fig. \ref{fig:step1}, are also seen
in Fig. \ref{fig:scatter}. The first current channels have moved to
around $z = 20\lambda_{e}$. At $z = 15 \lambda_{e}$, a second
current channel is created by the deflected jet electrons. This
periodic phenomenon involves the ions in a highly nonlinear process
but is beyond the scope of this Letter and will be explained in a
subsequent paper.

\section{Conclusions}
Using a three-dimensional relativistic particle-in-cell code, we
have investigated how an ambient magnetic field affects the dynamics
of a relativistic jet in the collisionless shock region. We have
examined how the different ambient magnetic topology and strength
affect the growth of the electron Weibel two-stream instability and
the associated electron acceleration. This instability is an
important mechanism in collisionless shocks.  It facilitates
momentum transfer between colliding plasma populations
\citep{bib:kazimura1998,bib:medvedevloeb,bib:nishikawa,bib:nishikawa2005,
bib:silva,bib:frederiksen} and can accelerate electrons to
nonthermal distributions \citep{bib:hoshino2002, bib:hededal2004,
bib:saito2003}. Collisionless shocks are found in the interface
between relativistic outflows (e.g., from gamma-ray bursts and
active galactic nuclei) and the surrounding medium (e.g., the ISM).

We find substantial differences between the cases of ambient
magnetic fields transverse and parallel to the jet flow. However,
common for both cases is that it takes an ambient magnetic field
strength much stronger than the strength of the magnetic field
typically found in the ISM to effectively suppress the Weibel
two-stream instability. In the case of a parallel magnetic field,
$\omega_{pe}/\omega_{ce}$ must be smaller than 5 to effectively
suppress the instability. This result is in good agreement with
two-dimensional simulations by \cite{bib:saito2003}, and thus this
limit seems independent of $\gamma_{jet}$. For a typical ISM density
of $10^6m^{-3}$, this corresponds to a milligauss magnetic field. We
emphasize the importance of $\omega_{pe}/\omega_{ce}$ as an
important parameter for collisionless shocks, as was also pointed
out by \cite{bib:shimada2004}.

In the case of perpendicular injection, the dynamics are different
from the parallel injection. Here, the electrons are deflected by
the magnetic field, and this creates a charge separation from the
less deflected ions. The charge separation drags the ambient and jet
electrons, and consequently they are strongly accelerated along the
$z$-direction. Furthermore, due to the Weibel instability, current
channels are generated around the ambient magnetic field, which is
curled and locally reversed.

These simulations provide insights into the complex dynamics of
relativistic jets. Further investigations are required to understand
the detailed physics involved. Larger simulations (above $10^9$
particles) with longer boxes are needed to cover the instability
domain of the ions, to investigate the full evolution of the
complicated dynamics, and to resolve the whole shock ramp.\\

This research (K.-I.N.) is partially supported by the National Science
Foundation awards ATM 9730230, ATM-9870072, ATM-0100997, and INT-9981508
and (C.B.H.) by a grant from the University of Copenhagen.
One of the authors C.B.H. would like to thank Dr. G. J. Fishman
for his support and hospitality during a stay at the
NSSTC - Huntsville Alabama, USA and Dr. \AA{}. Nordlund for his
help and guidance.


\begin{thebibliography}{99}

\bibitem[Frederiksen et al.(2004)]{bib:frederiksen}
Frederiksen, J. T., Hededal, C. B., Haugb\o lle, T., \& Nordlund, {\AA}.
2004, ApJ, 608, L13

\bibitem[Hededal et al.(2004)]{bib:hededal2004}
Hededal, C. B., Frederiksen, J. T., Haugb\o lle, T., \& Nordlund, {\AA}.
2004, ApJ, 617, L107


\bibitem[Hoshino \& Shimada(2002)]{bib:hoshino2002}
Hoshino, M., \& Shimada, N. 2002, ApJ, 572, 880

\bibitem[Kazimura et al.(1998)]{bib:kazimura1998}
Kazimura, Y., Sakai, J. I., Neubert, T.,  and Bulanov, S. V. 1998,
ApJ, 498, L183

\bibitem[Medvedev \& Loeb(1999)]{bib:medvedevloeb}
Medvedev, M. V., \& Loeb, A. 1999,
ApJ, 526, 697

\bibitem[Nishikawa et al.(2003)]{bib:nishikawa}
Nishikawa, K.-I.; Hardee, P.; Richardson, G.; Preece, R.; Sol, H.; Fishman,
G. J. 2003, ApJ, 595, 555

\bibitem[Nishikawa et al.(2005)]{bib:nishikawa2005}
Nishikawa, K.-I.; Hardee, P.; Richardson, G.; Preece, R.; Sol, H.;
Fishman, G. J.  2005, ApJ,622, 927

\bibitem[Sakai and Matsuo(2004)]{bib:sakai}
Sakai, J.-I. and Matsuo, A. 2004, Phys. Plasmas,  11, 3251

\bibitem[Saito \& Sakai(2003)]{bib:saito2003}
Saito, S., \& Sakai, J.I. 2003, Phys. Plasma, 11, 859

\bibitem[Shimada \& Hoshino(2004)]{bib:shimada2004}
Shimada, N., \& Hoshino, M. 2004, Phys. Plasmas, 11, 1840

\bibitem[Silva et at.(2003)]{bib:silva}
Silva, L.~O. and Fonseca, R.~A. and Tonge, J.~W. and Dawson, J.~M.
and Mori, W.~B. and Medvedev, M.~V. 2003, ApJ, 596, L121

\end{thebibliography}
\end{document}